\documentclass[reprint, aps, prl, floatfix, superscriptaddress]{revtex4-1}
\usepackage{graphicx}    
\usepackage{dcolumn}    
\usepackage{bm}             
\usepackage{amssymb}   
\usepackage{amsmath}    
\usepackage{amsthm}
\usepackage{mathrsfs}    
\usepackage{commath}   
\usepackage{subfigure}    
\usepackage{braket}
\usepackage{float}
\usepackage{color}
\usepackage{siunitx}
\usepackage[colorlinks=true, allcolors=blue]{hyperref}
\usepackage{hyphenat}

\newcommand{\beq}{\begin{equation}}
\newcommand{\eeq}{\end{equation}}

\begin{document}

\title{Heralded Generation and Detection of Entangled Microwave--Optical Photon Pairs}

\author{Changchun Zhong}
\email{zhong.changchun@yale.edu}
\affiliation{Department of Applied Physics, Yale University, New Haven, CT 06520, USA}
\affiliation{Yale Quantum Institute, Yale University, New Haven, CT 06520, USA}

\author{Zhixin Wang}
\affiliation{Department of Applied Physics, Yale University, New Haven, CT 06520, USA}
\affiliation{Yale Quantum Institute, Yale University, New Haven, CT 06520, USA}

\author{Changling Zou}
\affiliation{Key Laboratory of Quantum Information, CAS, University of Science and Technology of China, Hefei, Anhui 230026, China}

\author{Mengzhen Zhang}
\affiliation{Department of Applied Physics, Yale University, New Haven, CT 06520, USA}
\affiliation{Yale Quantum Institute, Yale University, New Haven, CT 06520, USA}

\author{Xu Han}
\affiliation{Yale Quantum Institute, Yale University, New Haven, CT 06520, USA}
\affiliation{Department of Electrical Engineering, Yale University, New Haven, CT 06520, USA}

\author{Wei Fu}
\affiliation{Yale Quantum Institute, Yale University, New Haven, CT 06520, USA}
\affiliation{Department of Electrical Engineering, Yale University, New Haven, CT 06520, USA}

\author{Mingrui Xu}
\affiliation{Yale Quantum Institute, Yale University, New Haven, CT 06520, USA}
\affiliation{Department of Electrical Engineering, Yale University, New Haven, CT 06520, USA}

\author{Shyam Shankar}
\affiliation{Department of Applied Physics, Yale University, New Haven, CT 06520, USA}
\affiliation{Yale Quantum Institute, Yale University, New Haven, CT 06520, USA}

\author{Michel H. Devoret}
\affiliation{Department of Applied Physics, Yale University, New Haven, CT 06520, USA}
\affiliation{Yale Quantum Institute, Yale University, New Haven, CT 06520, USA}

\author{Hong X. Tang}
\affiliation{Department of Applied Physics, Yale University, New Haven, CT 06520, USA}
\affiliation{Yale Quantum Institute, Yale University, New Haven, CT 06520, USA}
\affiliation{Department of Electrical Engineering, Yale University, New Haven, CT 06520, USA}

\author{Liang Jiang}
\email{liang.jiang@yale.edu}
\affiliation{Department of Applied Physics, Yale University, New Haven, CT 06520, USA}
\affiliation{Yale Quantum Institute, Yale University, New Haven, CT 06520, USA}

\date{\today}

\begin{abstract}
Quantum state transfer between microwave and optical frequencies is essential for connecting superconducting quantum circuits to coherent optical systems and extending microwave quantum networks over long distances. To build such a hybrid ``quantum Internet,'' an important experiment in the quantum regime is to entangle microwave and optical modes. Based on the model of a generic cavity electro-optomechanical system, we present a heralded scheme to generate entangled microwave--optical photon pairs, which can bypass the efficiency threshold for quantum channel capacity in direct transfer protocols. The preferable parameter regime for entanglement verification is identified. Our scheme is feasible given the latest experimental progress on electro-optomechanics, and can be potentially generalized to various physical systems.
\end{abstract}

\maketitle

The modular quantum architecture---moderate-sized quantum registers and memories connected by efficient communication channels---is a {competitive} approach toward a scalable quantum network \cite{Cirac1997,*Kimble2008,Jiang2007,Monroe2014}. Physically, it is comprised of natural or artificial ``atoms''---the nodes---and flying photons---the interconnects. As engineerable mesoscopic ``atoms,'' superconducting qubits \cite{LesHouches2003,Clarke2008} can strongly interact with microwave photons in cavities or waveguide resonators, known as the circuit quantum electrodynamics (cQED) architecture \cite{Blais2004,Girvin2011}. Lately, microwave photons have been employed to entangle remote transmon qubits \cite{Narla2016,Campagne-Ibarcq2018,Kurpiers2018} and cavity memories \cite{Axline2018,Chou2018}. However, the high loss in commercial microwave cables at room temperature prevents the transmission of quantum signals over long distances \cite{Kurpiers2017}. {In contrast,} optical photons stand out as quantum information carriers at large spatial scales---entanglement and teleportation have been demonstrated over kilometers through telecommunication fibers \cite{Tittel1998,Valivarthi2016} and over one thousand kilometers in free space \cite{Yin2017,*Ren2017}. Therefore, high-fidelity quantum state transfer between superconducting circuits and optical photons will greatly expand the quantum computing network, as well as bridging superconducting qubits and other quantum modules with coherent optical interfaces, {including} neutral atoms \cite{Matsukevich2004}, trapped ions \cite{Blinov2004}, defects in solids \cite{Togan2010,Koehl2011}, quantum dots \cite{Gao2012,DeGreve2012}, etc.

However, superconducting circuits do not have an optical transition. A quantum transducer is thus needed to interface microwave and optical photons in a quantum coherent manner. So far, most investigations of quantum transducers are based on direct quantum transduction \cite{Meng18}, which linearly converts input photons to output photons at different frequencies. Proposed direct microwave--optical (M--O) transducers involve cold alkali atoms \cite{Hafezi2012,Kiffner2016,Gard2017}, rare-earth-doped crystals \cite{Williamson2014,OBrien2014}, ferromagnetic magnons \cite{Hisatomi2016}, electro-optical devices \cite{Tsang2010,*Tsang2011,Javerzac-Galy2016}, and nanomechanical oscillators \cite{Andrews2014,Regal2011,Bochmann2013,Taylor2011,Barzanjeh2011,*Barzanjeh2012,Wang2012,Tian2010,*Tian2012,*Tian2014,Zou2016,Midolo2018,Bagci2014,Vainsencher16,Winger2011,Pitanti2015}. Recent experiments on cavity electro-optomechanics are very encouraging  \cite{Andrews2014,Higginbotham2018,Bagci2014,Bochmann2013,Vainsencher16,Han2014,Winger2011,Pitanti2015}, but many challenges remain. In these setups, microwave and mechanical resonators are coupled by either electrostatic \cite{Bagci2014,Andrews2014,Higginbotham2018} or piezoelectric forces \cite{Bochmann2013,Vainsencher16,Han2014}. Nanomembranes for electrostatic coupling usually vibrate at megahertz frequencies \cite{Bagci2014,Andrews2014,Higginbotham2018,Teufel2011,*Palomaki2013,*Reed2017}, resulting in a narrow conversion bandwidth and high added thermal noise. On the other hand, piezoelectric oscillators can be routinely fabricated with gigahertz frequencies \cite{O'Connell2010,Bochmann2013,Chu2017,*Chu2018,Han2014,Han2015,Han2016}, which couple to much lower thermal noise. Nevertheless, piezo-optomechanical converters also require large matched electromechanical and optomechanical cooperativities to achieve high conversion efficiency \cite{Han2016,Vainsencher16,Moritz18}. Although strong piezo-electromechanical coupling has been demonstrated with a cooperativity over 2000 \cite{Han2014,Zou2016}, it is still challenging to achieve a matched large optomechanical cooperativity in integrated piezo-optomechanical devices.

A direct photon converter is capable of transferring quantum states only if the conversion efficiency $\eta > 1/2$ \cite{eta}. In principle, we can bypass this stringent requirement by introducing two-way classical communication that first heralds successful entanglement generation \cite{Duan2001,Chou2005,Moehring2007,Lee2011,Hofmann2012,Bernien2013,Narla2016,Riedinger2018} and then completes the quantum state transfer by quantum teleportation \cite{Bennett1993,*Pirandola2015}. Quantum states can thus be transferred bi-directionally between microwave and optical frequencies, which connects superconducting quantum processors to an optical ``quantum Internet'' \cite{Kimble2008}. In this Letter, we propose such a heralded M--O entanglement generation and detection scheme, which is the first step of realizing this entanglement-based quantum transduction. We analyze its implementation in a generic cavity electro-optomechanical system with dissipation and thermal noise, and map {out} the preferable parameter ranges for manifesting M--O entanglement. Our scheme and analysis can be generalized to M--O transducers based on various physical platforms and therefore provide a new guideline for realizing coherent quantum state transfer between microwave and optical frequencies.

\begin{table*}[t]
\caption{Experimentally feasible parameters. Unless specified otherwise, these parameters apply to all figures and evaluations in the text.}\label{tab1}
\begin{center}
\begin{tabular}{c|c|c|c|c|c|c|c|c|c|c}
\hline
\hline
$g_{\text{em}}$/MHz  & $g_{\text{om,0}}$/kHz & $\kappa_ {\text{e,i}}$/kHz & $\kappa_{\text{o,i}}$/(GHz)  & $\kappa_ {\text{e,c}}$ & $\kappa_{\text{o,c}}$  & $\kappa_{\text{m}}$/kHz  &   $\bar{n}_\text{ba}$(T=1 K) & $\omega_{\text{m}}$/GHz & $\omega_{\text{e}}$/GHz & $\omega_{\text{o}}$/THz \\
\hline
$2\pi\times2.0$  & $2\pi\times 5.5$ & $2\pi\times 100$   &   $2\pi\times 0.24$   & $\sim$($\kappa_\text{e,i},10^3\kappa_ {\text{e,i}})$ & $\sim\kappa_{\text{o,i}}$  &  $2\pi\times20$   &    $\sim1.67$ &  $2\pi\times10$ &  $2\pi\times10$ &  $2\pi\times195$ \\
\hline
\hline
\end{tabular}
\end{center}
\label{default}
\end{table*}

\begin{figure*}[t]
\includegraphics[width=\textwidth]{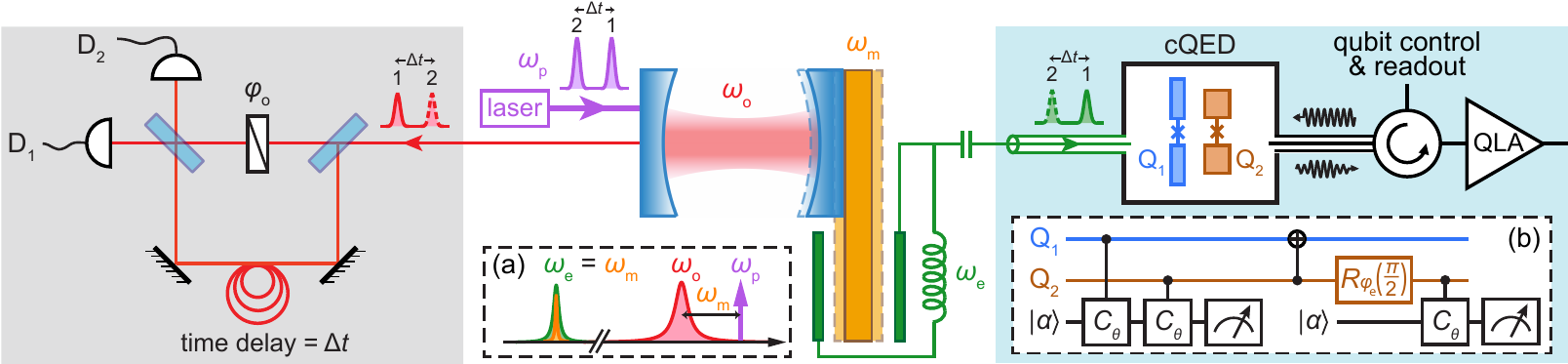}
\caption{Schematic setup for generating and detecting time-bin entanglement between optical and microwave photons. The thickness mode of a piezoelectric mechanical oscillator (yellow) is simultaneously coupled to a microwave $LC$ resonator (green) and an optical cavity (illustrated as a Fabry--P\'{e}rot interferometer). A pulsed laser (purple) pumps the optical cavity at its blue sideband. Depicted in inset (a) is the frequency landscape. In each generation round, two laser pulses are separated by $\Delta t$ in time. Optical photons (red) are analyzed by {a Franson-type} unbalanced Mach--Zehnder interferometer: on the long arm, the extended fiber delays the photon precisely by time $\Delta t$; on the short arm, photon phase is shifted by a variable $\varphi_\mathrm{o}$. Both beam splitters are 50/50. The outputs of the second beam splitter are sent into a pair of single-photon detectors ($\textrm{D}_1$ and $\textrm{D}_2$). On the microwave side, {photons (green) are analyzed in a circuit QED system consisting of two superconducting transmon qubits (blue and brown) with matched dispersive shifts to a cavity mode (black). High-fidelity single-shot readout is performed in reflection with the aid of a quantum-limited amplifier (QLA). Microwave photons are first converted to qubit excitations through stimulated Raman absorption.} Shown in inset (b) is the subsequent qubit control and readout operations---a joint parity measurement, a CNOT gate, and a single-qubit readout. The black line represents the readout cavity initialized in a coherent state $\ket{\alpha}$. A controlled-phase gate ($C_\theta$) {followed by a meter} represents the cQED dispersive readout. $R_{\varphi_\mathrm{e}}(\pi/2)$ stands for a $\pi/2$ rotation about the $(\sin\varphi_\mathrm{e},\cos\varphi_\mathrm{e},0)$ axis on the Bloch sphere. \label{fig_scheme}}
\end{figure*}

\textit{Entanglement generation.---}Without loss of generality, our physical discussions are based on the model depicted in Fig.~\ref{fig_scheme}. The thickness mode of a mechanical oscillator is on one side linearly coupled to a microwave resonator by piezoelectric force \cite{Zou2016, Han2016, Chu2017}, and on the other side parametrically coupled to an optical cavity by radiation pressure \cite{Aspelmeyer2014}. The frequencies of the optical, the mechanical, and the microwave resonators are at $\omega_{\mathrm{o}}$, $\omega_{\mathrm{m}}$, and $\omega_{\mathrm{e}}$, respectively. A laser at frequency $\omega_\mathrm{p}=\omega_\mathrm{o}+\Delta_\mathrm{p}$ pumps the optical cavity and populates it by $\bar{n}_\mathrm{o}$ photons on average. In the rotating frame of the pump, we write the linearized Hamiltonian of the system
\beq
\begin{split}
{\hat{H}}/{\hbar} &=  -\Delta_\mathrm{p} \hat{a}^\dagger\hat{a} + \omega_{\mathrm{m}} \hat{b}^\dagger \hat{b} + \omega_{\mathrm{e}} \hat{c}^\dagger \hat{c} - g_{\mathrm{em}} (\hat{b}^\dagger \hat{c} + \hat{b} \hat{c}^\dagger)\\
&\quad\, -g_{\mathrm{om,0}} \sqrt{\bar{n}_\mathrm{o}} (\hat{a}^\dagger + \hat{a}) (\hat{b}^\dagger + \hat{b}) ,
\end{split}
\eeq
where $\hat{a}$, $\hat{b}$, and $\hat{c}$ represent the optical, mechanical, and microwave modes; $g_{\mathrm{em}}$ and $g_{\mathrm{om,0}}$ are the piezoelectric and the single-photon optomechanical coupling rates.

To entangle microwave and optical photons, {as shown in the inset (a) of Fig.~\ref{fig_scheme},} we generate entangled phonon--photon pairs by driving an optomechanical parametric down-conversion process with a blue-sideband pump at $\Delta_\mathrm{p} = \omega_\mathrm{m}$ \cite{Lee2011, Riedinger2018, Riedinger2016, *Hong2017}. Meanwhile, mechanical excitations are swapped into the microwave resonator through the piezoelectric interaction. The M--O mode is thus approximately in a two-mode squeezed vacuum {$\ket{\psi_\mathrm{sq}(\lambda)}_\mathrm{oe} \simeq \sum_{N=0}^\infty \frac{\lambda^N}{N!} (\hat{a}^\dagger)^N(\hat{c}^\dagger)^N\ket{0}_\mathrm{o}\ket{0}_\mathrm{e}$}, in which subscripts ``o" and ``e" represent the optical and microwave modes. For a weak enough pump (the effective squeezing factor $\lambda\ll 1$), an output M--O photon pair can be generated through the coupling ports with probability $|\lambda|^2$.

{To use flying photons as quantum information carriers and demonstrate non-classical correlations, we can encode qubits {into} multiple modes with different polarizations \cite{Freeman1972, Aspect1981, *Aspect1982}, momenta \cite{Rarity1990}, time bins \cite{Marcikic2002}, frequency bins \cite{DTCE,*Olislager2010}, etc.} As an example, we present the scheme of generating time-bin entanglement in our proposed setup. As shown in Fig.~\ref{fig_scheme}, the optical cavity is pumped by two blue-sideband laser pulses separated by time $\Delta t$, which is within the coherence time of the pump laser and superconducting qubits (alternativelty, we can pump the cavity with continuous wave and select output by pulsed gates with time separation $\Delta t$). Denoting $\hat{a}^{(1,2)}_\text{out,c}$ and $\hat{c}^{(1,2)}_\text{out,c}$ as the optical and microwave output mode operators in time-bin 1 and 2, respectively, the M--O output modes is
\beq \label{eo_time_bin}
\begin{split}
{\ket{\Psi_\mathrm{tb}(\lambda)}_\mathrm{oe} } & \simeq \ket{0, 0}_\mathrm{o} \ket{0, 0}_\mathrm{e} + \lambda \hat{a}^{(1)\dagger}_\text{out,c}\hat{c}^{(1)\dagger}_\text{out,c}\ket{0, 0}_\mathrm{o} \ket{0, 0}_\mathrm{e} \\
& \quad + \lambda \hat{a}^{(2)\dagger}_\text{out,c}\hat{c}^{(2)\dagger}_\text{out,c}\ket{0, 0}_\mathrm{o} \ket{0, 0}_\mathrm{e} + O(\lambda^2),
\end{split}
\eeq
where for either mode, $\ket{0,0}$ denotes the state with zero photon in the first and second time bin. Neglecting the $O(\lambda^2)$ terms and discarding the zero-photon events by postselection, we obtain a time-bin Bell state $\frac{\sqrt{2}}{{2}} (\hat{a}^{(1)\dagger}_\text{out,c}\hat{c}^{(1)\dagger}_\text{out,c} + \hat{a}^{(2)\dagger}_\text{out,c}\hat{c}^{(2)\dagger}_\text{out,c})\ket{0, 0}_\mathrm{o} \ket{0, 0}_\mathrm{e} $ {with probability $|\lambda|^2$}.

\textit{Flying photon measurement.---}Optical time-bin qubits can be detected with an unbalanced Mach--Zehnder interferometer \cite{Franson1989}. As depicted on the left of Fig.~\ref{fig_scheme} (the gray panel), given the relative time delay between the long and short arms precisely matching $\Delta t$, a photon in the first time bin passing through the long arm and a photon in the second time bin passing through the short arm will meet and interfere at the second beam splitter. A click at $\mathrm{D}_{1,2}$ projects the optical qubit on $\ket{\varphi_\mathrm{o}^\pm}_\mathrm{o}=\frac{1}{\sqrt{2}}(\ket{1, 0}_\mathrm{o} \pm e^{i\varphi_\mathrm{o}}\ket{0, 1}_\mathrm{o})$, in which $\varphi_\mathrm{o}$ is an adjustable phase shift. Note that the unconditional maximum efficiency of this projective readout is 0.5, because half of the click events correspond to a photon in the first time bin passing through the short arm or a photon in the second time bin passing through the long arm, and thus produce no interference \cite{Tittel2000}. {Maximum visibility of unity} can be recovered by discarding these early and late counts through postselection.

On the microwave side (the blue panel on the right of Fig.~\ref{fig_scheme}), flying photons in two time bins are first converted to excitations of two transmon qubits in a cQED system through stimulated two-photon Raman absorption \cite{Campagne-Ibarcq2018}, after which the transmons and the optical modes are entangled as $\frac{1}{\sqrt{2}} (\ket{1, 0}_\mathrm{o} \ket{eg} + \ket{0, 1}_\mathrm{o} \ket{ge} )$, where $g$ ($e$) denotes the ground (first excited) state of a transmon.
{After a joint parity measurement (to be explained in the next paragraph),} a CNOT gate \cite{Paik2016} followed by a high-fidelity single-shot qubit readout \cite{Hatridge2013} projects the two transmons on $\frac{1}{\sqrt{2}}(\ket{eg}\pm e^{-i\varphi_\mathrm{e}}\ket{ge})$, where $\varphi_e$ can be continuously adjusted from 0 to $2\pi$. {We thus effectively project the microwave photons on $\ket{\varphi_\mathrm{e}^\pm}_\mathrm{e}=\frac{1}{\sqrt{2}}(\ket{1,0}_\mathrm{e} \pm e^{-i\varphi_\mathrm{e}} \ket{0,1}_\mathrm{e})$.}

During the experiment, heralded signals that indicate successful entanglement generation events are produced by measuring the output photons in both microwave and optical domains---an M--O Bell state contains one and only one photon pair. Specifically, on the microwave side, instead of counting photons, we can measure the parity of the two-qubit state immediately after the stimulated Raman absorption and postselect on the condition that only one transmon is excited \cite{Riste2013, Shankar2013, *Liu2016}. Excluded by heralding are null events $\ket{0, 0}_\mathrm{e} \ket{0, 0}_\mathrm{o}$, undecayed higher-order generation events, and the cases where the flying photons are lost in the paths. Although not increasing the generation rate, this heralded scheme significantly improves the fidelities of the obtained Bell pairs and enables the entanglement verification given {non-negligible transmission loss and limited detection efficiencies.}

\begin{figure}[t]
\includegraphics[width=\columnwidth]{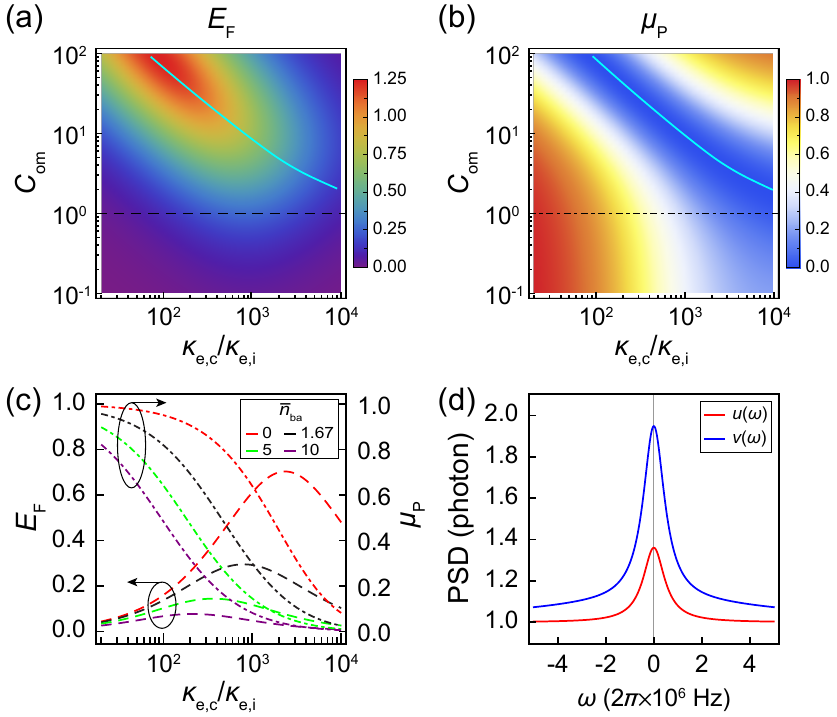}
\caption{M--O output state. Parameters in Tab.\ref{tab1} apply to all figures. (a) Entanglement of formation $E_\text{F}$ and (b) state purity $\mu_\text{P}$ versus ${C}_{\mathrm{om}}$ and the ratio $\kappa_\text{e,c}/\kappa_\text{e,i}$. The cyan lines trace the parameters such that $C_\text{om}-1=C_\text{em}$ is satisfied. (c) $E_\text{F}$ and $\mu_\text{P}$ plotted with varied thermal noises and fixed $C_\text{om}=1$---the black dashed lines in (a) and (b). (d) The optical (in red) and microwave (in blue) output power spectrum densities, where ${C}_\text{om}=1,\kappa_\text{e,c}/\kappa_\text{e,i}=150$ are used. \label{fig2}}
\end{figure}

\textit{Dissipation and thermal noise.---}In reality, the ideal squeezed vacuum $\ket{\psi_\mathrm{sq}(\lambda)}_\mathrm{oe}$ generated in the parametric down-conversion process is degraded to a generic two-mode Gaussian state by the dissipation and thermal fluctuations of the modes. We label the optical, microwave, and mechanical dissipation rates by $\kappa_\mathrm{o}= \kappa_\mathrm{o,i}+\kappa_\mathrm{o,c}$, $\kappa_\mathrm{e}=\kappa_\mathrm{e,i}+\kappa_\mathrm{e,c}$, and $\kappa_\mathrm{m}$ (the subscript `i' for internal loss, `c' for external coupling), and assume {the thermal phonon (photon) population of the mechanical (microwave) {dissipation bath is $\bar{n}_\mathrm{ba} = (e^{\hbar \omega_\mathrm{m(e)}/k_\mathrm{B} T}-1)^{-1}$} on average, while the optical resonator and the optical and microwave coupling ports are purely subject to vacuum fluctuations.} These assumptions comply with the experimental conditions: {in reality, we might need to vary the temperature of the electro-optomechanical device from 10 mK to a few kelvins for higher mechanical quality factor and better power handling capability. Meanwhile, the gigahertz microwave mode can be radiatively cooled close to its ground state if it's very overcoupled to the 10 mK bath ($\kappa_\mathrm{e,c}\gg \kappa_\mathrm{e,i}$).

The output two-mode Gaussian state can be derived analytically in the frequency domain combining the Heisenberg-Langevin equations of motion with the input-output theory. Denoting the output M--O state quadratures as $\textbf{x}_\text{oe}^\text{out} = \{\hat{q}_\text{o}, \hat{p}_\text{o}, \hat{q}_\text{e}, \hat{p}_\text{e}\}$, we write down the corresponding covariance matrix in the standard form
\beq\label{vou}
\mathbf{V}_{\mathrm{oe}}^\mathrm{out}=
\begin{pmatrix}
u(\omega) & 0 & -w(\omega) & 0\\
0 & u(\omega) & 0 & w(\omega)\\
-w(\omega) &0 & v(\omega) & 0\\
0 & w(\omega) & 0 & v(\omega)
\end{pmatrix},
\eeq
whose matrix elements reflect the corresponding quadrature correlations ({see Ref.~\cite{suppl} for details}). To characterize the output M--O state, we show its power spectrum densities $u(\omega)=\braket{|\hat{q}_\text{o}(\omega)|^2}$ (optical mode) and $v(\omega)=\braket{|\hat{q}_\text{e}(\omega)|^2}$ (microwave mode) in Fig. \ref{fig2}(d). The microwave mode has higher spectrum density due to its intrinsic coupling to the thermal bath. Also, we numerically calculate the entanglement of formation $E_\mathrm{F}$ ({see Ref.~\cite{suppl} for the definition) of the output state as a function of the optomechanical cooperativity $C_{\mathrm{om}}=4g^2_\text{om}/\kappa_\text{o}\kappa_\text{m}$ and the microwave readout ratio $\kappa_\text{e,c}/\kappa_\text{e,i}$, which can be tuned in experiments via the optical pump power and the position of the microwave readout probe, respectively. Using feasible parameters listed in Tab. \ref{tab1}, $E_\text{F}$ with resonant frequency ($\omega=0$ in rotating frame) is plotted in Fig. \ref{fig2}(a), which effectively demonstrates the entanglement of the output modes. $E_\mathrm{F}>0$ indicates the continuous-variable entanglement under any finite two-mode squeezing and $E_\mathrm{F}$ reaches its maximum when the electromechanical cooperativity $C_{\mathrm{em}}=4g^2_\text{em}/\kappa_\text{e}\kappa_\text{m} \simeq C_{\mathrm{om}}-1$ where strong parametric down conversion (PDC) dominates. In these regimes, the system tends to be unstable and generates entangled state with extremely low state purity $\mu_\text{P}$, as shown in Fig. \ref{fig2}(b) {(see Ref. \cite{pu} for purity definition)}. In Fig. \ref{fig2}(c), the $E_\text{F}$ and $\mu_\text{P}$ are shown with $C_\text{om}=1$ and different thermal nosies. As expected, higher thermal noise decreases the entanglement and the state purity. It is worth pointing out that in these plots, the parameter regime $g_\text{em}<\kappa_\text{e,c}$ is chosen in order to avoid electromechanical strong coupling and mode splitting, which can potentially complicate the microwave photon detection and thus not preferred in the entanglement generation. Nevertheless, entanglement in the strong coupling regime would be interesting for further investigations.

\begin{figure}[t]
\includegraphics[width=\columnwidth]{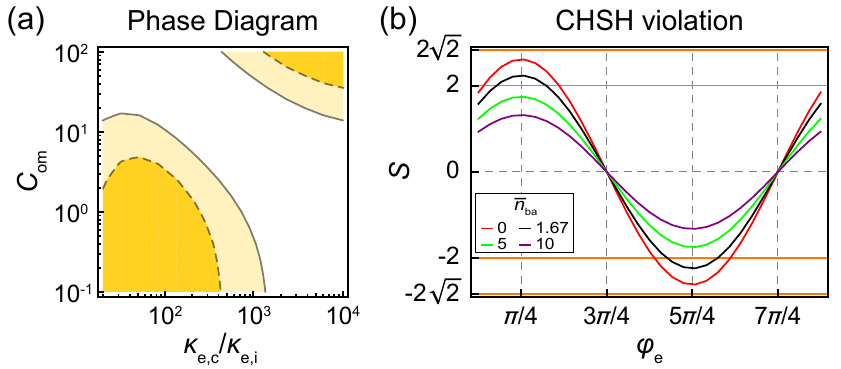}
\caption{(a) ``Phase diagram'' of Bell state fidelity and max CHSH inequality violation for ${\bar{n}_\mathrm{ba}}=1.67$. Solid and dashed curves represent the parameter threshold contours of $F_\mathrm{lb}=1/2$ and $|S|_\mathrm{max}=2$, respectively. The regions with $F_\text{lb}>1/2$ ($|S|_\mathrm{max}>2$) are shaded with light yellow (orange). (b) $S(0,\varphi_\mathrm{e};\pi/2,\varphi_\mathrm{e}+\pi/2)$ for varied thermal noises, given $C_{\mathrm{om}}=1$ and $\kappa_\text{e,c}/\kappa_\text{e,i}=150$. In these calculations, the measurement efficiencies are assumed to be 1/2 on both the microwave and optical sides. \label{fig3}}
\end{figure}

\textit{Entanglement verification.---}A lower bound of the entanglement fidelity of the M--O output state with respect to the ideal Bell state $\frac{\sqrt{2}}{2} (\hat{a}^{(1)\dagger}_\text{out,c}\hat{c}^{(1)\dagger}_\text{out,c} + \hat{a}^{(2)\dagger}_\text{out,c}\hat{c}^{(2)\dagger}_\text{out,c} )\ket{\text{vac}}$ is given by \cite{Blinov2004,Bernien2013}
\beq
F_\mathrm{lb} = \sum_{\substack{\nu=\pm \\ \varphi=0,\frac{\pi}{2}}}\! \! \frac{p^{\nu \nu}_{\varphi \varphi}}{2} - \frac{p^{+-}_{\frac{\pi}{2}\frac{\pi}{2}}}{2} - \frac{p^{-+}_{\frac{\pi}{2}\frac{\pi}{2}}}{2} - \sqrt{p^{+-}_{00} p^{-+}_{00}},
\eeq
in which $p^{\mu \nu}_{\varphi_\mathrm{o} \varphi_\mathrm{e}}$ is the \emph{normalized} probability of detecting $\ket{\varphi_\mathrm{o}^\mu}_\mathrm{o}$ and $\ket{\varphi_\mathrm{e}^\nu}_\mathrm{e}$ on the optical and microwave sides in a Bell state measurement. $F_\mathrm{lb}>1/2$ strictly indicates the M--O entanglement. Such regions are delineated in Fig. \ref{fig3}(a), which shows that a better entanglement fidelity should avoid the strong PDC parameter regime. Although $E_\text{F}$ is large at these regimes, the highly-mixed-entangled state (shown in Fig. \ref{fig2}(b)) is not suitable for this entanglement verification. This is also consistent with the previous discussions that weakly squeezed M--O output states better approximate the two-mode squeezed vacuum and thus the ideal time-bin Bell state. 

Furthermore, a stronger entanglement manifestation that excludes local hidden-variables is the violation of the Clauser--Horne--Shimony--Holt (CHSH) inequality \cite{Clauser1969}, which can be tested in our proposed setup by measuring the correlation quantity
\beq
\begin{split}
\label{sexpr}
S(\varphi_\mathrm{o}, \varphi_\mathrm{e}; \varphi_\mathrm{o}^\prime, \varphi_\mathrm{e}^\prime) &\equiv \\
E(\varphi_\mathrm{o}, \varphi_\mathrm{e}) + E&(\varphi_\mathrm{o}^\prime, \varphi_\mathrm{e}^\prime) + E(\varphi_\mathrm{o}^\prime, \varphi_\mathrm{e}) - E(\varphi_\mathrm{o}, \varphi_\mathrm{e}^\prime),
\end{split}
\eeq
in which {$E(\varphi_\mathrm{o}, \varphi_\mathrm{e}) \equiv p^{++}_{\varphi_\mathrm{o} \varphi_\mathrm{e}} + p^{--}_{\varphi_\mathrm{o} \varphi_\mathrm{e}} - p^{+-}_{\varphi_\mathrm{o} \varphi_\mathrm{e}} - p^{-+}_{\varphi_\mathrm{o} \varphi_\mathrm{e}}$} can be acquired by detecting photonic and superconducting qubits on the $\{\ket{\varphi_\mathrm{o}^\pm}_\mathrm{o}, \ket{\varphi_\mathrm{e}^\pm}_\mathrm{q}\}$ basis. Choosing $\varphi_\mathrm{o}^\prime-\varphi_\mathrm{o} = \varphi_\mathrm{e}^\prime-\varphi_\mathrm{e}=\pi/2$, we simulate the typical curves of $S$ in Fig.~\ref{fig3}(b) for varied thermal baths. It can be seen that $S$ sinusoidally depends on the phase variable and reaches its maximum when $\varphi_\mathrm{e}-\varphi_\mathrm{o} = \pi/4 + k\pi$ ($k \in \mathbb{N}$). For low bath photon numbers, clear inequality violation ($|S|_\mathrm{max}>2$) is observed. As the thermal noise increases, the violation is gradually destroyed. The $|S|_\mathrm{max}>2$ regions are delineated in Fig.~\ref{fig3}(a), which shows that the threshold for CHSH violation is more stringent than $F_\mathrm{lb}>1/2$.

\textit{Discussion.---}The coincidence detection rate can be written as $R_\text{c}=R_\text{ac}+R_\text{cc}$ \cite{OCAQ}, where the first and the second parts denote the accidental and the correlated counts, respectively. Ideally, they are approximated by $R_\text{ac}=R_\text{o}R_\text{e}\tau_b$ and $R_\text{cc}=\int_0^{\tau_b} \abs{R_\text{oe}}^2 d\tau$, where $\tau_b$ is the photon collection time window, $R_\text{o}=\braket{\hat{a}^{\dagger}_\text{out,c}(t)\hat{a}_\text{out,c}(t)}$ and $R_\text{e}=\braket{\hat{c}^{\dagger}_\text{out,c}(t)\hat{c}_\text{out,c}(t)}$ are the optical and microwave photon generation rates, respectively, and $R_\text{oe}=\braket{\hat{a}^\dagger_\text{out,c}(t)\hat{c}^\dagger_\text{out,c}(t+\tau)}$ is the photon correlations, which reaches maximum when the detector time delay $\tau\simeq0$. Considering a typical experiment repetition time $\sim 10\ \si{\us}$ \cite{Campagne-Ibarcq2018} that includes $1\ \si{\us}$ Raman absorption (determining the detector time window) and using $C_\text{om}=1$ and $\kappa_\text{e,c}/\kappa_\text{e,i}=150$, the joint detection rate is found approximately on the order of $10^4\ \si{Hz}$. 

In reality, the detection scheme suffers from photon losses (filter or transmission losses), detector inefficiency, and dark counts. Denoting $T_\text{o},D_\text{o},\eta_\text{o}$ ($T_\text{e},D_\text{e},\eta_\text{e}$) as the optical (microwave) photon transmissivity, dark count rate and detector efficiency, respectively, the accidental and correlated counting rates are replaced by $R_\text{ac}=(\eta_\text{o}T_\text{o}R_\text{o}+D_\text{o})(\eta_\text{e}T_\text{e}R_\text{e}+D_\text{e})\tau_\text{b}$ and $R_\text{cc}=\eta_\text{o}\eta_\text{e}T_\text{o}T_\text{e}\int_0^{\tau_b} \abs{R_\text{oe}}^2 d\tau$, which shows that the joint detection rates become lower due to the transmission loss and detector inefficiency, while dark counts increase the accidental counts. To ensure a good entanglement fidelity, $R_\text{cc}\gg R_\text{ac}$ is required, which leads to $g^{(2)}\gg 2+\xi_\text{o}+\xi_\text{e}+\xi_\text{o}\xi_\text{e}$. $g^{(2)}=R_\text{c}/R_\text{ac}$ is the second order correlation function \cite{PDCP}, and $\xi_\text{o}=\frac{D_\text{o}}{\eta_\text{o}T_\text{o}R_\text{o}}$ and $\xi_\text{e}=\frac{D_\text{e}}{\eta_\text{e}T_\text{e}R_\text{e}}$ are the ratios of dark counts to the signal counts. As expected, suppressing detector dark counts are beneficial for identifying entanglement. Meanwhile, the detector efficiencies, the transmissivities, and the detector time window are essential for observing the M--O entanglement in a reasonable amount of time, and should be simultaneously optimized in the experimental design.

Our proposal is compatible with recent experiments on cavity electro-optomechanics. Given a relatively smaller $C_{\mathrm{om}}$ \cite{Han2014,Zou2016,Vainsencher16}, the system can be well prepared in the cooperativity mismatched regime, leading to efficient entangled photon pair generation and detection even in the presence of a few thermal noise photons. Furthermore, although our analysis in this Letter is grounded on electro-optomechanics, the entanglement generation and detection scheme can be potentially generalized to other types of M--O converters based on parametric interactions---for instance, the electro-optic converters \cite{Rueda2016,Fan2018}---and thus shed light on M--O quantum state transfer in various physical systems.

\begin{acknowledgements}
We acknowledge insightful discussions with Chiao-Hsuan Wang, Yiwen Chu, Vijay Jain. We acknowledge support from the ARL-CDQI (W911NF-15-2-0067, W911NF-18-2-0237), ARO (W911NF-18-1-0020, W911NF-18-1-0212), ARO MURI (W911NF-16-1-0349 ), AFOSR MURI (FA9550-14-1-0052, FA9550-15-1-0015), DOE (DE-SC0019406), NSF (EFMA-1640959), and the Packard Foundation. 
\end{acknowledgements}

\bibliographystyle{apsrev4-1}
\bibliography{M2O_theory_bib}

\begin{thebibliography}{95}
\bibitem [{pu()}]{pu}%
  \BibitemOpen
  \href@noop {} {}\bibinfo {note} {For a given state $\hat{\rho}$, its purity
  is defined as $\mu_\text{P}=$tr$(\hat{\rho}^2)$. The purity equals to one for
  a pure output state.}\BibitemShut {Stop}%
\end{thebibliography}

\end{document}


\title{Supplemental Material for ``Heralded Generation and Detection of Entangled Microwave--Optical Photon Pairs''}

\author{Changchun Zhong}
\email{zhong.changchun@yale.edu}
\affiliation{Department of Applied Physics, Yale University, New Haven, CT 06520, USA}
\affiliation{Yale Quantum Institute, Yale University, New Haven, CT 06520, USA}

\author{Zhixin Wang}
\affiliation{Department of Applied Physics, Yale University, New Haven, CT 06520, USA}
\affiliation{Yale Quantum Institute, Yale University, New Haven, CT 06520, USA}

\author{Mengzhen Zhang}
\affiliation{Department of Applied Physics, Yale University, New Haven, CT 06520, USA}
\affiliation{Yale Quantum Institute, Yale University, New Haven, CT 06520, USA}

\author{Changling Zou}
\affiliation{Key Laboratory of Quantum Information, CAS, University of Science and Technology of China, Hefei, Anhui 230026, China}

\author{Xu Han}
\affiliation{Yale Quantum Institute, Yale University, New Haven, CT 06520, USA}
\affiliation{Department of Electrical Engineering, Yale University, New Haven, CT 06520, USA}

\author{Wei Fu}
\affiliation{Yale Quantum Institute, Yale University, New Haven, CT 06520, USA}
\affiliation{Department of Electrical Engineering, Yale University, New Haven, CT 06520, USA}

\author{Mingrui Xu}
\affiliation{Yale Quantum Institute, Yale University, New Haven, CT 06520, USA}
\affiliation{Department of Electrical Engineering, Yale University, New Haven, CT 06520, USA}

\author{Shyam Shankar}
\affiliation{Department of Applied Physics, Yale University, New Haven, CT 06520, USA}
\affiliation{Yale Quantum Institute, Yale University, New Haven, CT 06520, USA}

\author{Michel H. Devoret}
\affiliation{Department of Applied Physics, Yale University, New Haven, CT 06520, USA}
\affiliation{Yale Quantum Institute, Yale University, New Haven, CT 06520, USA}

\author{Hong X. Tang}
\affiliation{Department of Applied Physics, Yale University, New Haven, CT 06520, USA}
\affiliation{Yale Quantum Institute, Yale University, New Haven, CT 06520, USA}
\affiliation{Department of Electrical Engineering, Yale University, New Haven, CT 06520, USA}

\author{Liang Jiang}
\email{liang.jiang@yale.edu}
\affiliation{Department of Applied Physics, Yale University, New Haven, CT 06520, USA}
\affiliation{Yale Quantum Institute, Yale University, New Haven, CT 06520, USA}

\date{\today}

\maketitle

\beginsup

\section{Output M--O state}\label{op_state}
This section derives the covariance matrix of the output M--O state in the main text. Starting from Eq.~(1)---the linearized electro-optomechanical Hamiltonian in the rotating frame of the pump laser, we write down the Heisenberg-Langevin equations and input--output relations
\begin{align}
\dot{\mathbf{a}} &= \mathbf{M} \, \mathbf{a} + \mathbf{N} \, \mathbf{a}_\mathrm{in},\\
\mathbf{a}_{\mathrm{out}} &= \mathbf{N}^\mathrm{T} \mathbf{a} - \mathbf{a}_{\mathrm{in}},
\end{align}
in which $\mathbf{a}$ is a vector which collects all the mode operators and similarly $\mathbf{a}_\mathrm{in}$ and $\mathbf{a}_\mathrm{out}$ collect all the input and output mode operators
\begin{align}
\mathbf{a} &= \left( \hat{a}, \, \hat{a}^\dagger, \, \hat{c},\hat{c}^\dagger, \, \hat{b}, \, \hat{b}^\dagger \right)^\mathrm{T}, \\
\mathbf{a}_{\mathrm{in}} &= \left(\hat{a}_{\mathrm{in,c}}, \, \hat{a}^\dagger_{\mathrm{in,c}}, \, \hat{a}_{\mathrm{in,i}}, \, \hat{a}^\dagger_{\mathrm{in,i}}, \, \hat{c}_{\mathrm{in,c}}, \, \hat{c}^\dagger_{\mathrm{in,c}}, \, \hat{c}_{\mathrm{in,i}}, \, \hat{c}^\dagger_{\mathrm{in,i}}, \, \hat{b}_{\mathrm{in}}, \, \hat{b}^\dagger_{\mathrm{in}}\right)^\mathrm{T},\\
\mathbf{a}_\mathrm{out} &= \left(\hat{a}_\mathrm{out,c}, \, \hat{a}^\dagger_\mathrm{out,c}, \, \hat{a}_\mathrm{out,i}, \, \hat{a}^\dagger_\mathrm{out,i}, \, \hat{c}_\mathrm{out,c}, \, \hat{c}^\dagger_\mathrm{out,c}, \, \hat{c}_{\mathrm{out,i}}, \, \hat{c}^\dagger_{\mathrm{out,i}}, \, \hat{b}_{\mathrm{out}}, \, \hat{b}^\dagger_{\mathrm{out}}\right)^\mathrm{T},\\
\mathbf{M} &=
\begin{pmatrix}
i\Delta_{\mathrm{p}}-\frac{\kappa_{\mathrm{o}}}{2}	&	 0	&	0	&	0	&	ig_\mathrm{om}	&	ig_\mathrm{om}\\
0	&	-i\Delta_{\mathrm{p}}-\frac{\kappa_{\mathrm{o}}}{2}	&	0	&	0	&	-ig_\mathrm{om}		&	-ig_\mathrm{om} \\	
0	&	0	&	-i\omega_{\mathrm{e}}-\frac{\kappa_{\mathrm{e}}}{2}	&	0	&	ig_\mathrm{em}		&	0 \\
0	&	0	&	0	&	i\omega_{\mathrm{e}}-\frac{\kappa_{\mathrm{e}}}{2}	&	0	&	-ig_\mathrm{em} \\
ig_\mathrm{om}	&	ig_\mathrm{om}	&	ig_\mathrm{em}	&	0	&	-i\omega_{\mathrm{m}}-\frac{\kappa_{\mathrm{m}}}{2}		&	0  \\
-ig_\mathrm{om}	&	-ig_\mathrm{om}	&	0	&	-ig_\mathrm{em} 	&	0	&	i\omega_{\mathrm{m}}-\frac{\kappa_{\mathrm{m}}}{2} \\
\end{pmatrix}, \\
\mathbf{N} &=
\begin{pmatrix}
\sqrt{\kappa_{\mathrm{o,c}}}	&	 0	&	\sqrt{\kappa_{\mathrm{o,i}}}	&	0	&	0	&	0	&	0	&	0 	&	0	&	0 \\
0	&	 \sqrt{\kappa_{\mathrm{o,c}}}	&	0	&	\sqrt{\kappa_{\mathrm{o,i}}}	&	0	&	0	&	0	&	0 	&	0	&	0 \\
0	&	 0	&	0	&	0	&	\sqrt{\kappa_{\mathrm{e,c}}}	&	0	&	\sqrt{\kappa_{\mathrm{e,i}}}	&	0 	&	0	&	0 \\
0	&	 0	&	0	&	0	&	0	&	\sqrt{\kappa_{\mathrm{e,c}}}	&	0	&	\sqrt{\kappa_{\mathrm{e,i}}} 	&	0	&	0 \\
0	&	 0	&	0	&	0	&	0	&	0	&	0	&	0 	&	\sqrt{\kappa_{\mathrm{m}}}	&	0 \\
0	&	 0	&	0	&	0	&	0	&	0	&	0	&	0 	&	0	&	\sqrt{\kappa_{\mathrm{m}}} \\
\end{pmatrix}.
\end{align} 
Transform the mode operators into the frequency domain,
\begin{align}
\mathbf{a}[\omega] &= \int^\infty_{-\infty} \mathbf{a}(t) e^{i\omega t} \: \mathrm{d}t,\\
\mathbf{a}_\mathrm{in}[\omega] &= \int^\infty_{-\infty} \mathbf{a}_\mathrm{in}(t) e^{i\omega t} \: \mathrm{d}t,\\
\mathbf{a}_\mathrm{out}[\omega] &= \int^\infty_{-\infty} \mathbf{a}_\mathrm{out}(t) e^{i\omega t} \: \mathrm{d}t.
\end{align}
$\mathbf{a}_\mathrm{in}[\omega]$ and $\mathbf{a}_\mathrm{out}[\omega]$ can be linked through a scattering matrix $\mathbf{S}_a[\omega]$,
\beq
\mathbf{a}_\mathrm{out}[\omega] = \mathbf{S}_a[\omega] \, \mathbf{a}_\mathrm{in}[\omega] = \left[ \mathbf{N}^\mathrm{T} \left(-i \omega \mathbf{D}_6-\mathbf{M} \right)^{-1} \mathbf{N} - \mathbf{I}_{10} \right] \mathbf{a}_\mathrm{in}[\omega],
\eeq
where $\mathbf{I}_{10}$ denotes the 10-dimensional identity matrix, and $\mathbf{D}_6=\mathrm{diag}(1, \, -1,\,1,\,-1,\,1,\,-1)$. 
\\

Define the quadrature operators in the form of
\begin{equation}
\begin{pmatrix}
\hat{q}^a \\ \hat{p}^a
\end{pmatrix}
=
\begin{pmatrix}
1&1\\
-i&i
\end{pmatrix}
\begin{pmatrix}
\hat{a}\\
\hat{a}^\dagger
\end{pmatrix}.
\end{equation}
Consequently, the input and output quadrature fields,
\beq
\mathbf{x}_{\mathrm{in}} = \left(\hat{q}^a_{\mathrm{in,c}}, \, \hat{p}^a_{\mathrm{in,c}}, \, \hat{q}^a_{\mathrm{in,i}}, \, \hat{p}^a_{\mathrm{in,i}}, \, \hat{q}^c_{\mathrm{in,c}}, \, \hat{p}^c_{\mathrm{in,c}}, \, \hat{q}^c_{\mathrm{in,i}}, \, \hat{p}^c_{\mathrm{in,i}}, \, \hat{q}^b_{\mathrm{in}}, \, \hat{p}^b_{\mathrm{in}}\right)^\mathrm{T},
\eeq 
and 
\beq
\mathbf{x}_{\mathrm{out}} = \left(\hat{q}^a_\mathrm{out,c}, \, \hat{p}^a_\mathrm{out,c}, \, \hat{q}^a_\mathrm{out,i}, \, \hat{p}^a_\mathrm{out,i}, \, \hat{q}^c_\mathrm{out,c}, \, \hat{p}^c_\mathrm{out,c}, \, \hat{q}^c_{\mathrm{out,i}}, \, \hat{p}^c_{\mathrm{out,i}}, \, \hat{q}^b_{\mathrm{out}}, \, \hat{p}^b_{\mathrm{out}}\right)^\mathrm{T},
\eeq are linked by a symplectic mapping, which preserves the Gaussian properties of the fields,
\beq\label{s8}
\mathbf{x}_\mathrm{out}[\omega] = \mathbf{S}_x[\omega] \, \mathbf{x}_\mathrm{in}[\omega] = \mathbf{Q} \, \mathbf{S}_a[\omega] \, \mathbf{Q}^{-1} \mathbf{x}_\mathrm{in}[\omega],
\eeq
Here
\beq
\mathbf{Q}=\mathbf{I}_{5}\otimes\begin{pmatrix}
1&1\\
-i&i
\end{pmatrix}.
\eeq
\\

A Gaussian quantum state can be fully characterized by its first and second quadrature moments---the mean value $\braket{\mathbf{x}}$ and the covariance matrix $\mathbf{V}$ with elements $V_{ij}=\frac{1}{2} \left\langle \{\hat{x}_i-\braket{\hat{x}_i},\hat{x}_j-\braket{\hat{x}_j} \} \right \rangle$, given $\mathbf{x}=(\hat{x}_1, \, \hat{x}_2, \, ... \,)^\mathrm{T}$ \cite{Weedbrook2012}. After the Gaussian unitary channel,
\begin{align}
\left\langle \mathbf{x}_\mathrm{out} [\omega] \right \rangle  &= \mathbf{S}_x [\omega]  \left\langle \mathbf{x}_\mathrm{in} [\omega] \right \rangle ,\\
\mathbf{V}_\mathrm{out}[\omega] &=\mathbf{S}_x[\omega]  \, \mathbf{V}_\mathrm{in} [\omega]  \, \mathbf{S}_x^\mathrm{T} [\omega] .
\end{align}
In our proposed experimental setting, we assume the mechanical and electrical dissipation bath share a non-negligible thermal population $\bar{n}_{\mathrm{ba}}$, while all the other input and output ports are only subject to vacuum fluctuations. We therefore write 
\beq
\mathbf{V}_\mathrm{in}=\mathrm{diag}[\, \mathbf{I}_6, \, (2\bar{n}_\mathrm{ba}+1)\,\mathbf{I}_4 \,].
\eeq 
After applying the rotating wave approximation, we obtain the $4 \times 4$ reduced covariance matrix in the standard form between the output optical and electrical fields $\mathbf{x}_\text{oe}^\text{out}=\{\hat{q}_\mathrm{o}^\mathrm{out}, \hat{p}_\mathrm{o}^\mathrm{out}, \hat{q}_\mathrm{e}^\mathrm{out}, \hat{p}_\mathrm{e}^\mathrm{out}\} \equiv \{\hat{q}^a_\mathrm{out,c}[-\omega], \hat{p}^a_\mathrm{out,c}[-\omega], \, \hat{q}^c_\mathrm{out,c}[\omega], \, \hat{p}^c_\mathrm{out,c}[\omega] \}$
\beq\label{vout}
\mathbf{V}_\mathrm{oe}^\mathrm{out}
=
\begin{pmatrix}
\mathbf{V}_u & \mathbf{V}_w \\
\mathbf{V}_w & \mathbf{V}_v
\end{pmatrix}
=
\begin{pmatrix}
u(\omega) & 0 & -w(\omega) & 0 \\
0 & u(\omega) & 0 & w(\omega)\\
-w(\omega) & 0 & v(\omega) & 0\\
0 & w(\omega) & 0 & v(\omega)
\end{pmatrix}.
\eeq
$u(\omega)$ and $v(\omega)$ give the power spectrum distribution in the main text. When the frequency is right on resonance ($\omega=0$ in the rotating frame), the matrix elements take the following simple form
\beq\label{melement}
\begin{split}
u &= 1 + \frac{8 \zeta_\mathrm{o} \mathcal{C}_{\mathrm{om}} \left[1 + \mathcal{C}_{\mathrm{em}} + \bar{n}_\mathrm{ba} + \mathcal{C}_{\mathrm{em}} \bar{n}_\mathrm{ba} (1 -\zeta_\mathrm{e}) \right]} {\left( 1 + \mathcal{C}_{\mathrm{em}} - \mathcal{C}_{\mathrm{om}}\right)^2},\\
v &= 1 + \frac{8 \zeta_\mathrm{e} \left[ \mathcal{C}_{\mathrm{em}} (\mathcal{C}_{\mathrm{om}} + \bar{n}_\mathrm{ba}) + (\mathcal{C}_{\mathrm{om}}-1)^2 \bar{n}_\mathrm{ba} (1 -\zeta_\mathrm{e}) \right]} {\left( 1 + \mathcal{C}_{\mathrm{em}} - \mathcal{C}_{\mathrm{om}} \right)^2},\\
w &= \frac{4 \sqrt{\zeta_\mathrm{o} \zeta_\mathrm{e} \mathcal{C}_{\mathrm{em}} \mathcal{C}_{\mathrm{om}}} \left\{1 + \mathcal{C}_{\mathrm{em}} + \mathcal{C}_{\mathrm{om}} + 2 \bar{n}_\mathrm{ba} \left[ \zeta_\mathrm{e}+\mathcal{C}_\mathrm{om} (1 - \zeta_\mathrm{e}) \right] \right\}} {\left( 1 + \mathcal{C}_{\mathrm{em}} - \mathcal{C}_{\mathrm{om}} \right)^2},
\end{split}
\eeq
with $\zeta_\mathrm{o}=\kappa_{\mathrm{o,c}}/\kappa_{\mathrm{o}}$, $\zeta_\mathrm{e}=\kappa_{\mathrm{e,c}}/\kappa_{\mathrm{e}}$, and $\mathcal{C}_{\mathrm{om}}\equiv{4 g_{\mathrm{om}}^2}/{\kappa_{\mathrm{o}}\kappa_{\mathrm{m}}}$ and $\mathcal{C}_{\mathrm{em}}\equiv{4g_{\mathrm{em}}^2}/{\kappa_{\mathrm{e}}\kappa_{\mathrm{m}}}$. $1 + \mathcal{C}_{\mathrm{em}} - \mathcal{C}_{\mathrm{om}}=0$ makes the matrix elements diverge, which corresponds to the strong parametric down conversion regime.  
\\

\section{Entanglement of formation}
Entanglement of formation ($E_\text{F}$) of a mixed bipartite state is defined as the infimum of the average Von Neumann entropy taken over all its possible pure state decompositions. It was proved to be an effective entanglement measure for Gaussian states \cite{Marian2008,*Tserkis2017}. For a general two mode Gaussian state, a lower bound is given by the formula
\beq
E_\mathrm{F} = \cosh^2r \, \log_2\! \left(\cosh^2 r \right) - \sinh^2r \, \log_2\! \left(\sinh^2 r \right),
\eeq
where $r$ is the minimum amount of anti-squeezing needed to disentangled the state
\beq
r = \frac{1}{4} \ln \left(\frac{\gamma-\sqrt{\gamma^2-\beta_+\beta_-}}{\beta_-} \right),
\eeq
with
\begin{align}
\gamma &= 2 \left(\det\mathbf{V}^\mathrm{out}_\mathrm{oe} + 1 \right)-(u(\omega)-v(\omega))^2,\\
\beta_{\pm} &= \det \mathbf{V}_u + \det \mathbf{V}_v-2\det \mathbf{V}_w + 2 u(\omega) v(\omega) + 2 w^2(\omega) \pm 4 w(\omega) (u(\omega)+v(\omega)).
\end{align} 
The lower bound is saturated for a two mode Gaussian state in the standard form. In the main text, we used the above formula to plot the $E_\text{F}$ in the weak coupling regime (take $\omega=0$) and the strong coupling regime (take $\omega=g_\text{em}$).

\section{Coincidence counting probability}

The probability of jointly detecting an optical photon at time $t$ and a microwave photon at $t+\tau$ is proportional to the Glauber formula \cite{TQTO,*CAIS,ETPW} $P_\text{c}\propto \braket{\hat{a}_{\text{out,c}}^\dagger(t)\hat{c}_{\text{out,c}}^\dagger(t+\tau)\hat{c}_{\text{out,c}}(t+\tau)\hat{a}_{\text{out,c}}(t)}$, where $\tau$ is the time delay between the two photon detectors. This formula assumes the detectors can resolve single photon, which in general is hard to achieve. To analyze our experimental scheme, we model both photon detectors on the optical side and the circuit QED system on the microwave side as on/off detectors without photon-number resolution. Such a detector can be described by the positive operator-valued measure \cite{Ferraro2005}
\begin{align}
\hat{\Pi}_{\mathrm{off}} &= \sum_{N=0}^\infty (1-\eta)^N \ket{N} \bra{N},\\
\hat{\Pi}_{\mathrm{on}} &= \mathbf{I} - \hat{\Pi}_{\mathrm{off}},
\end{align}
in which $\eta$ is the detector efficiency. Thus, the joint detection probability should be replaced by
\beq
P_\text{c}\propto \braket{\hat{\Pi}^\text{o}_\text{on}(t)\otimes\hat{\Pi}_\text{on}^\text{e}(t+\tau) }.
\eeq
To give a theoretical estimation, we now focus on the case of $\tau=0$. Then the joint detection probability can be expressed as a frequency integral
\beq
\begin{split}
P_\text{c}\propto \int_{-\infty}^\infty P_\text{on}^\text{o,e}(\omega)d\omega,
\end{split}
\eeq
where we denote $P_\text{on}^\text{o,e}(\omega)=\text{tr}(\hat{\rho}\hat{\Pi}^\text{o}_\text{on}(\omega)\otimes\hat{\Pi}_\text{on}^\text{e}(\omega))$ as the coincidence counting probability for frequency $\omega$. $\hat{\rho}=\ket{\psi}\bra{\psi}$ is the state density matrix.

Given any Gaussian state with Wigner function
\beq\label{wig}
W(\textbf{x})=\frac{\exp(-\frac{1}{2}\textbf{x}^T\textbf{V}^{-1}\textbf{x})}{(2\pi)^2\sqrt{\text{det}\textbf{V}}},
\eeq
where $\mathbf{V}$ is the covariance matrix. we can accordingly calculate the ``click'' probability
\beq
P_\mathrm{on} = \int W(\mathbf{x}) \tilde{\Pi}_\mathrm{on}(\mathbf{x}) \: \mathrm{d} \mathbf{x},
\eeq
in which $\tilde{\Pi}_\mathrm{on}(\mathbf{x})$ is the Weyl transform of $\hat{\Pi}_{\mathrm{on}}$,
\beq
\tilde{\Pi}_\mathrm{on} (\mathbf{x}) = \tilde{\Pi}_\mathrm{on} (\mathbf{q}, \mathbf{p}) = \int \left \langle \mathbf{q} + \frac{\mathbf{q}^\prime}{2} \right | \hat{\Pi}_\mathrm{on} \left | \mathbf{q}-\frac{\mathbf{q}^\prime}{2} \right \rangle e^{i \mathbf{p}^\mathrm{T} \mathbf{q}^\prime} \: \mathrm{d} \mathbf{q}^\prime.
\eeq
Here we have grouped $\mathbf{x}$ into $\mathbf{q}=(q_1, \, q_2, \, \dots)^\mathrm{T}$ and $\mathbf{p}=(p_1, \, p_2, \, \dots)^\mathrm{T}$. Thus the coincidence counting probability between two detectors is
\beq \label{prob_c}
P_\mathrm{on}^\text{o,e}(\omega) = \int W(\mathbf{x}) \tilde{\Pi}_\mathrm{on}^\text{o,e}(\mathbf{x}) \: \mathrm{d} \mathbf{x},
\eeq
in which
\beq \label{weyl}
\tilde{\Pi}_\mathrm{on}^\text{o,e} (\mathbf{x}) = \int \left \langle \mathbf{q} + \frac{\mathbf{q}^\prime}{2} \right | \hat{\Pi}^\text{o}_\mathrm{on}\otimes \hat{\Pi}^\text{e}_\mathrm{on} \left | \mathbf{q}-\frac{\mathbf{q}^\prime}{2} \right \rangle e^{i \mathbf{p}^\mathrm{T} \mathbf{q}^\prime} \: \mathrm{d} \mathbf{q}^\prime.
\eeq
\\

We now apply this formulation to analyze the detection of time-bin entanglement. On the optical side, the Franson interferometer is effectively a 50/50 beam splitter transforming the output time-bin modes from the electro-optomechanical device to the measurement basis $\ket{\varphi_\mathrm{o}^\pm}$,
\beq
\begin{pmatrix}
\hat{a}_{\varphi_\mathrm{o}}^+ \\
\hat{a}_{\varphi_\mathrm{o}}^- 
\end{pmatrix}
= \frac{1}{\sqrt{2}}
\begin{pmatrix}
1 & e^{-i \varphi_\mathrm{o}} \\
-e^{i \varphi_\mathrm{o}}& 1
\end{pmatrix}
\begin{pmatrix}
\hat{a}_\mathrm{out,c}^{(1)} \\
\hat{a}_\mathrm{out,c}^{(2)}
\end{pmatrix}
,
\eeq
where the two time bins are labeled by superscripts (1) and (2). Similarly, on the electrical side,
\beq
\begin{pmatrix}
\hat{c}_{\varphi_\mathrm{e}}^+ \\
\hat{c}_{\varphi_\mathrm{e}}^- 
\end{pmatrix}
= \frac{1}{\sqrt{2}}
\begin{pmatrix}
1 & e^{i \varphi_\mathrm{e}} \\
-e^{-i \varphi_\mathrm{e}}& 1
\end{pmatrix}
\begin{pmatrix}
\hat{c}_\mathrm{out,c}^{(1)} \\
\hat{c}_\mathrm{out,c}^{(2)}
\end{pmatrix}
.
\eeq
Following the procedure in Sec.~\ref{op_state} of this Supplemental Material, we can derive the $8 \times 8$ covariance matrix of quadrature operators $\{ \hat{q}_{\varphi_\mathrm{o}}^+ , \, \hat{p}_{\varphi_\mathrm{o}}^+ , \, \hat{q}_{\varphi_\mathrm{o}}^-, \, \hat{p}_{\varphi_\mathrm{o}}^- , \, \hat{q}_{\varphi_\mathrm{e}}^+ , \, \hat{p}_{\varphi_\mathrm{e}}^+ , \, \hat{q}_{\varphi_\mathrm{e}}^-, \, \hat{p}_{\varphi_\mathrm{e}}^- \}$. The reduced $4 \times 4$ covariance matrix of $\{ \hat{q}_{\varphi_\mathrm{o}}^\mu , \, \hat{p}_{\varphi_\mathrm{o}}^\mu , \, \hat{q}_{\varphi_\mathrm{e}}^\nu , \, \hat{p}_{\varphi_\mathrm{e}}^\nu \}$ ($\mu, \nu = \pm$),
\beq
\mathbf{V}(\varphi_\mathrm{o},\varphi_\mathrm{e}; \mu, \nu) =
\begin{pmatrix}
\mathbf{V}_a & \mathbf{V}_{ac}\\
\mathbf{V}_{ca} & \mathbf{V}_{c}
\end{pmatrix},
\eeq
contains the information about the \emph{raw} coincidence counting probability between $\ket{\varphi_\mathrm{o}^\mu}$ and $\ket{\varphi_\mathrm{e}^\nu}$. Each submatrix is $2 \times 2$. Given the efficiencies of the single-photon optical and microwave ``detectors'' to be $\eta_\mathrm{o}$ and $\eta_\mathrm{e}$ respectively, using Eqs.~(\ref{wig}), (\ref{prob_c}), and (\ref{weyl}), we obtain
\beq\label{coin}
P_{(\varphi_\mathrm{o},\varphi_\mathrm{e}; \mu, \nu)}(\omega)=1-\frac{2}{(2-\eta_{\mathrm{o}})\sqrt{\mathrm{det}\bm{\Sigma}_{a}}}-\frac{2}{(2-\eta_{\mathrm{e}})\sqrt{\mathrm{det}\bm{\Sigma}_{c}}}+\frac{4}{(2-\eta_{\mathrm{o}})(2-\eta_{\mathrm{e}})\sqrt{\mathrm{det}\bm{\Sigma}_{ac}}},
\eeq
in which 
\begin{align}
\mathbf{\Sigma}_{a} &= \frac{\eta_{\mathrm{o}}}{2-\eta_{\mathrm{o}}} \mathbf{V}_{a} + \mathbf{I}_2, \\
\mathbf{\Sigma}_{c} &= \frac{\eta_{\mathrm{e}}}{2-\eta_{\mathrm{e}}} \mathbf{V}_{c} + \mathbf{I}_2, \\
\mathbf{\Sigma}_{ac} &= \mathbf{V} \cdot \left( \frac{\eta_{\mathrm{o}}}{2-\eta_{\mathrm{o}}}\mathbf{I}_2 \oplus \frac{\eta_{\mathrm{e}}}{2-\eta_{\mathrm{e}}}\mathbf{I}_2 \right) + \mathbf{I}_4.
\end{align}
According to Eq.~(\ref{vout}),
\beq
\begin{split}
P_{(\varphi_\mathrm{o},\varphi_\mathrm{e}; \mu, \nu)}(\omega) =& 1 - \frac{2}{2+\eta_{\mathrm{o}}(u(\omega)-1)} - \frac{2}{2+\eta_{\mathrm{e}}(v(\omega)-1)}\\
& + \frac{4}{[2+\eta_{\mathrm{o}}(u(\omega)-1)][2+\eta_{\mathrm{e}}(v(\omega)-1)] - \eta_{\mathrm{o}}\eta_{\mathrm{e}}w^2(\omega)[1 +  \mu\nu\cos(\varphi_\mathrm{o}-\varphi_\mathrm{e})]}.
 \end{split}
\eeq
Thus the total joint detection rate can be obtained by integrate all frequency contributions,
\beq
\begin{split}
P_\text{c}(\varphi_\mathrm{o},\varphi_\mathrm{e}; \mu, \nu) = \int_{-\infty}^\infty P_{(\varphi_\mathrm{o},\varphi_\mathrm{e}; \mu, \nu)}(\omega) d\omega.
\end{split}
\eeq

\section{Entanglement fidelity lower bound, Bell inequality}

In the main text, $p^{\mu \nu}_{\varphi_\mathrm{o} \varphi_\mathrm{e}}$ ($\mu, \nu = \pm$) denotes the \emph{normalized} probability of detecting $\ket{\varphi_\mathrm{o}^\mu}_\mathrm{o}$ and $\ket{\varphi_\mathrm{e}^\nu}_\mathrm{e}$ in a Bell state measurement,
\beq \label{puv}
p^{\mu \nu}_{\varphi_\mathrm{o} \varphi_\mathrm{e}} = \frac{P_\text{c}(\varphi_\mathrm{o},\varphi_\mathrm{e}; \mu, \nu)} {P_\text{c}(\varphi_\mathrm{o},\varphi_\mathrm{e}; +,+) + P_\text{c}(\varphi_\mathrm{o},\varphi_\mathrm{e}; +, -) + P_\text{c}(\varphi_\mathrm{o},\varphi_\mathrm{e}; -, +) + P_\text{c}(\varphi_\mathrm{o},\varphi_\mathrm{e}; -,-)}.
\eeq
Physically, this normalization procedure corresponds to postselecting coincidence counting events based on the heralding signals. Using Eq.~(\ref{puv}), we numerically calculated the entanglement fidelity lower bound $F_\text{lb}$ (Eq. (4) in the main text) and the quantity $S$ in Bell inequality (Eq. (5) in the main text). 

We also give the analytical expressions for $F_\text{lb}$ and $S$ with on resonance condition $\omega=0$, which corresponds to using a very narrow-bandwidth filter, and could analytically capture the main physical pictures. We obtain
\beq
F_\mathrm{lb} = \frac{\mathscr{A} - \mathscr{B}}{2 - \frac{4}{2 + \eta_\mathrm{o}(u-1)} - \frac{4}{2 + \eta_\mathrm{e} (v-1)} + \mathscr{A} + \mathscr{B}},
\eeq
in which $\mathscr{A}= \frac{4}{[2 + \eta_{\mathrm{o}} (u-1)] [2 + \eta_\mathrm{e} (v-1)] - \eta_\mathrm{o} \eta_\mathrm{e} w^2 },
\mathscr{B} = \frac{4}{[2 + \eta_{\mathrm{o}} (u-1)] [2 + \eta_\mathrm{e} (v-1)]}$, and
\beq 
S\left(0, \varphi_\mathrm{e}; \frac{\pi}{2}, \varphi_\mathrm{e} + \frac{\pi}{2} \right) = \frac{2(\mathscr{E}-\mathscr{F})}{2-\frac{4}{2+\eta_{\mathrm{o}}(u-1)}-\frac{4}{2+\eta_{\mathrm{e}}(v-1)}+\mathscr{E}+\mathscr{F}} -
\frac{2(\mathscr{G} - \mathscr{H})}{2-\frac{4}{2 + \eta_{\mathrm{o}}(u-1)} - \frac{4}{2 + \eta_{\mathrm{e}}(v-1)} + \mathscr{G} + \mathscr{H}},
\eeq
in which 
\[
\begin{split}
\mathscr{E} &= \frac{4}{[2 + \eta_\mathrm{o} (u-1)][2 + \eta_{\mathrm{e}}(v-1)] - \eta_\mathrm{o}\eta_\mathrm{e} w^2 \cos^2(\pi/4 - \varphi_\mathrm{e}/2)}, \\
\mathscr{F} &= \frac{4}{[2 + \eta_\mathrm{o} (u-1)][2 + \eta_{\mathrm{e}}(v-1)] - \eta_\mathrm{o}\eta_\mathrm{e} w^2 \cos^2(\pi/4 + \varphi_\mathrm{e}/2)}, \\
\mathscr{G} &= \frac{4}{[2 + \eta_\mathrm{o} (u-1)][2 + \eta_{\mathrm{e}}(v-1)] - \eta_\mathrm{o}\eta_\mathrm{e} w^2 \cos^2(\varphi_\mathrm{e}/2)}, \\
\mathscr{H} &= \frac{4}{[2 + \eta_\mathrm{o} (u-1)][2 + \eta_{\mathrm{e}}(v-1)] - \eta_\mathrm{o}\eta_\mathrm{e} w^2 \sin^2(\varphi_\mathrm{e}/2)}.
\end{split}
\]
Particularly,
\beq
|S|_\mathrm{max} = \left|S \left(0, \frac{\pi}{4} + k \pi; \frac{\pi}{2}, \frac{3\pi}{4} + k \pi \right) \right| = \left|\frac{4(\mathscr{C}-\mathscr{D})}{2-\frac{4}{2 + \eta_{\mathrm{o}}(u-1)}-\frac{4}{2 + \eta_{\mathrm{e}}(v-1)} + \mathscr{C} + \mathscr{D}} \right|,
\eeq
in which
\begin{align}
\mathscr{C} &= \frac{4}{[2 + \eta_\mathrm{o} (u-1)][2 + \eta_\mathrm{e}(v-1)] - \eta_\mathrm{o}\eta_\mathrm{e} w^2 \cos^2(\pi/8)},\\
\mathscr{D} &= \frac{4}{[2 + \eta_\mathrm{o} (u-1)][2 + \eta_\mathrm{e}(v-1)] - \eta_\mathrm{o}\eta_\mathrm{e} w^2 \sin^2(\pi/8)}.
\end{align}

\bibliographystyle{apsrev4-1}
\bibliography{M2O_sup_bib}